\documentclass[12pt,preprint]{aastex}

\slugcomment{Submitted to \apj}

\shorttitle{Search for enhancements of muons from Solar flares and GRBs}
\shortauthors{Navia et al.}

\begin{document}

\title{Is there an enhancement of muons at sea level from transient events?}


\author{C. E. Navia, C. R. A. Augusto, M. B. Robba, M. Malheiro, and H. Shigueoka}
\affil{Instituto de F\'{i}sica, Universidade Federal Fluminense, 24210-130 Niter\'{o}i, RJ, Brazil}

\email{navia@if.uff.br}



\begin{abstract}

In a recent study of a search for enhancements from the galactic center with muons
at sea level using the TUPI muon telescope, we have found several
ground level enhancements (GLEs) as
very sharp peaks above  the count rate background. This paper reports  a consistent 
analysis of two  GLEs observed in December 2003 and detected after an up-grade of 
the data acquisition system, which includes a noise filter and which 
allows us to verify that the GLEs are not mere background fluctuations.
The main target of this study is a search for the origin of the GLEs.
The results show that one of them has a strong correlation with a solar flare,
while the other has an unknown origin, because there is neither a satellite report 
of a solar flare, nor prompt X-ray emission,  and nor a excess of nuclei
during the raster scan where the GLE was observed.
Even so, two possibilities are analyzed: the solar flare hypothesis and the
gamma ray burst (GRB) hypothesis. We show, by using the FLUKA Monte Carlo results
for photo-production, that under certain conditions there is the possibility of an 
enhancement of muons at sea level from GeV GRBs. 

\end{abstract}




\section{Introduction}

In gamma-ray astronomy, an important energy band of the electromagnetic 
spectrum, the gap region between gamma-rays up to 30 GeV covered by 
satellites and the region starting from $\sim 0.2$ TeV covered by  ground-based 
experiments is still not very well explored. In order to reach this
energy region, since January of 2002 the TUPI muon telescope, located 
in Niter\'{o}i, Brazil, has been in operation. 
Due to a limited aperture ($9.5^0$ of opening angle), the TUPI telescope     
is on the boundary between telescopes with a very small field of view, like the air     
Cherenkov telescopes, and the small air shower arrays, characterized     
by a large field of view.

On the other hand, a muon telescope at ground level can be used as an astronomical 
telescope even if the muon's parent particles are neutral, such as photons. 
Muons are expected as  products of 
cosmic ray (protons, nuclei) interaction with the nuclei of the 
atmosphere via $\pi^{\pm}$ production followed by the decay 
$\pi^{\pm} \to \mu^{\pm} \nu_{\mu}$.
However, a small fraction of muons have their origin in photonuclear
reactions induced by gamma rays. 
The ability to distinguish these "photonuclear"
muons from background muons depends upon statistics, the strength and 
energy spectrum of the emitting source and some characteristic of the 
experiment such as the angular resolution. 

In order to increase the sensitivity of TUPI telescope, an up-grade of the acquisition 
system was made on December, 2003 and included a new 
(on-line) discrimination level to increase
the counting rate background from approximately one per 20 second to approximately 18 
per 10 second. In addition, as the experiment uses scintillator plastics and due to other limitations
(see section 3) the counting rate has a dependence with the (on-line) pulse-height 
discrimination and this new data acquisition allows us to build a noise filter.
The aim of this framework is to make a consistent analysis of ground level events,   
observed as very sharp peaks above the count rate background and found during a 
study of a possible muon excess from the galactic center. 

In this article we report two GLEs detected on December 2003 and obtained under 
these new conditions. We give emphasis to the study of their origin, including the 
temporal characteristics on the basis of the time profile shapes of the GLEs and duration.
We show, that the first GLE has a strong correlation with solar flares. Because the second
GLE has an unknown origin, two possibilities are analyzed.
The paper is organized as follows: In section 2, we present the TUPI telescope characteristics.
In section 3, the raster search technique is presented. The analysis of the two GLEs is
presented in section 4. In section 5 the local effects are analyzed. In section 6 the possible 
existence of a window to observe GRBs at the ground is discussed and finally in section 7 we 
present our conclusions.

\section{The TUPI telescope}

The TUPI muon telescope is installed on the campus of the
Universidade Federal Fluminense, Niter\'{o}i, Rio de Janeiro-Brazil.
The position is: latitude: $22^0 54'33''$ S,
longitude: $43^0 08' 39''$ W, at sea level. 
The telescope is inside a building, under two flagstones 
of concrete as is shown in Fig.1. The flagstones's thickness has a dependence on the 
zenith angle ($150\;g/cm^2$ in average). For large zenith angles ($\theta > 76^0$) and in 
the directions South and West  there are two walls with a thickness close to $165\;g/cm^2$ 
and behind the walls we have open sea. However, for the North and East directions, besides 
the two walls, there are the buildings of Niteroi City. Consequently, the North
and East directions are 
opaque for almost horizontal atmospheric muons.  The flagstones or walls   
increase the muon energy threshold. The telescope can detect muons  at sea level with  
energies greater than the $\sim 0.3$ GeV required to penetrate the two flagstones.  
In addition the flagstones contribute to the $e/\mu$   
separation, consequently reducing the noise due to non-muon particles.   

Most of the particles ($\sim 95\%\;at\; E>1 GeV$)  observed at sea level 
are muons. This means that the telescope
computes mainly the muon intensity in the atmosphere initiated by cosmic
rays (mainly protons), giving the coincidence counting rate  of
multi-elements (plastic scintillator detectors with $0.5m\times
0.5m \times 35 cm$). Each plastic scintillator 
is viewed by a 5 cm diameter Hamamatsu RS21 photomultiplier. 
The unit has a energy threshold of $\sim 10 MeV$ and  
is fed by very stable power supplies.
A servo-mechanism (with an equatorial assembly) 
allows that the axis of the telescope can be pointed so as to 
accompany a given source.
Details of the experimental setup of the TUPI muon telescope as well as a 
study of the muon background characteristics have previously been reported  
\citep{augusto03}.

In our experiment, only particles close to the telescope 
axis are detected ($65.3 cm^2sr$ of aperture and $9.5^0$  of opening angle).
Essentially shower particles coming from other directions are rejected by a 
combination of coincidence between the two telescope detectors and anti-coincidence 
with another detector off the telescope axis 
constituting a trigger system as is shown in Fig.2. 

The relative efficiency for the detection of charged particles by plastic scintillator 
in relation to NaI(Tl) crystal is approximately 50\%. The plastic scintillator light 
emission does not reach the optimum frequency region of the photomultiplier, the de-excitation and 
light emission do not always occurs preferentially via low-lying impurity levels, and the
wavelength of maximum emission happens in a band of waves. 
In addition the efficiency is also compromised because the coupling between the scintillator and the 
photocathode (light guide) in our experiment is air and there is the possibility of reflections  
in the walls of the container where each unit of detection is mounted.
Consequently two identical charged particles can give different light-pulse amplitude.
In our experiment, a pulse-height discriminator (on line) is used to select only those pulses 
above a certain amplitude. This eliminates electronic noise and reduces the background (random) 
count rate of the system. The counting rate (even of single muons) has a
dependence on the pulse-height discrimination (on-line) used in the experiment.
In order to see the counting rate dependence on the pulse height discrimination,
now every time that the trigger 
conditions are satisfied, the height of the pulses in all the detectors 
are read and stored in a file. This allows us to choose several (off-line) pulse-height
discriminator levels via software starting from a certain value.

The data acquisition is made on the on the basis of the
Advantech PCI-1711/73 card with a A/D conversion up to 1000 kHz sampling rate.
All the steps from signal discrimination to the coincidence and anti-coincidence 
are made via software, using the virtual instrument technique. The application 
programs were written using the Lab-View tools.

\section{Raster search technique}

The observation of the galactic center began on 2003 June 26,   
and consisted of on-source and off-source raster scan across parallel lines 
in declination during a sidereal half-day (12 hours). 
This is approximately the time that the galactic center is above our horizon in 
every 24 hours. Starting from 2003 December, the counter rate every 10 seconds was used to observe 
a possible muon excess in the direction of the galactic center  
($Declination=-29^0,\;Right Ascension=17^h42^m$).

The atmospheric muon flux originating from the decay of charged pions and kaons  produced by cosmic 
rays in the atmosphere  is nearly isotropic. However, at lower energy (sub-GeV to GeV) the muon flux 
is influenced by the magnetic field of the Earth \citep{hayakawa69}.
Consequently the muon angular distributions  are
quite different for different sites on the globe.
There are two the main geomagnetic effects on the muon flux observed at ground level:
a) The East-West effect, the muon flux is highest (lowest) for direction coming from the West (East).
The effect is a consequence of the fact that
the electric charge of the primary cosmic ray is mainly positive.
The muon flux is a convolution of the primary spectrum and carries the imprint of this geomagnetic effect.
b) The azimuthal dependence on the positive-to-negative ratio of muons, a considerable amount of the negative
excess is observed for muons coming from the East. The effect is a consequence of the geomagnetic 
deflection being different for positive and negative particles, as well as the dependence of the 
path length of a muon on azimuth; a positive muon coming from the East has a longer path length
than a negative one.   
These geomagnetic effects distort the zenith angle distribution of sub-GeV to GeV muons
during a raster scan, because the measures (galactic center) begin around the East direction
and they finish around West direction.

Fig.3 summarizes the situation for the raster scan on $2003/12/22$. In the upper part, the telescope 
output, raw data ($counting\;rate\;vs\;universal\;time$)
is shown. In the lower part the squares represent the integral ($\mu^+ +\mu^-$) flux
without any corrections, and this is evaluated 
from the raw data each hour. One hour of scan corresponds to $15^0$ in hour angle and this also corresponds 
approximately to the zenith angle, when the declination is not very large, (i.e $-29^0$).
In spite of the telescope's relatively low angular resolution ($9.5^0)$ and
the small statistics in the counting rate of muons during one hour, inside the angular bins of
$15^0$, it is possible to observe the characteristics indicated above    
in the muons flux during a raster scan, such as, for example, a muon flux excess from the West direction.  

A quantitative result of the West-East effect during a raster scan can be obtained from two symmetrical 
points, those with the some zenith angle, with one pointing to the West and the other one to the East.
Table 1 shows some TUPI results, where the W-E asymmetry is defined as the ratio $[(W-E)/(W+E)]$. This relative
value is free from the detection bias and normalization. In Table 1 the Okayama results 
\citep{tokiwa03} are also included for comparison.


On the other hand, measurements of the absolute muon intensities by telescopes require 
measurements of the muon energy. In the present stage of our experiment, the detector 
is only a directional muon counter telescope and only 
relative intensities can be obtained.
The obtaining of absolute muon flux requires normalization to available absolute measurements.
Even so, it is well known that below 1 GeV there is a systematic
dependence on location due to Earth's magnetic field (geomagnetic effect).  In addition,
ground observations using new generations of spectrometers such as the CAPRICE 
\citep{boezio00,kremer99} (in 1997) and the
BESS \citep{tanizaki03}(in 2001), both at Mt. Summer ($892\;g/cm^2$) have show that the 
difference in the muon 
intensity is as large as about 20\% around 1 GeV. The difference can be attributed to the 
effect of solar modulation. An annual variation of the muon flux at sea level 
also has been reported, with changes by about 5\% at 1 GeV  expected.

Here, we have opted for the following strategy: measurements of the integral muon 
intensities has been made at axis orientations of $0^0$ up to $84.5^0$. 
In order to reduce the  geomagnetic effects,  the measurements are 
made at fixed azimuth angle (the telescope's axis always pointed southerly), 
and a fit
with the AMH (Texas A $\&$ M-University of Houston) function \citep{green79},  
after a numerical integration and extrapolation to 0.3 GeV, has been made. 
Fig.4 summarizes the situation. We can see that the Tupi muon intensities are systematically 
a little higher than the AMH fit for large zenith angle. However, the TUPI muon intensities
are consistent with the AMH fit. The several sources of modulation 
of sub-GeV to GeV muons can explain these differences.

\section{Ground level events}

The two GLEs reported in this paper were detected on 2003 December,
under new conditions of (on-line) pulse-height discrimination level.
The duration of the two GLEs is  big enough so 
that the GLEs cannot be interpreted as a surviving solitary small shower, 
or remains of an extensive shower both begun by conventional cosmic rays.
Besides this, the chances of detection of showers with the telescope are small, because
they are rejected.

\subsection{The first ground level event (GLE 2003/12/02)}

The GLE 2003/12/02 was detected during a raster scan off-source. 
The light curve shape
presents several peaks, as shown in Fig.4 (lower panel), with two peaks being  dominant, 
at the beginning and at the end of the GLE. The excesses for these two peaks can be seen over a 
background rate of approximately 18 coincidences per 10 seconds with significance levels of
$9.7\sigma$ and $10.5\sigma$ respectively. 

The analysis began with a search for the origin of the GLE.
Following Fig.4 it is possible to see the similarity
between the light curve of the two last solar flares on 2003 December 2, reported by the 
GOES satellite \citep{combs03}
\footnote{(Combs 2003)is available at http://sec.noaa.gov/ftpmenu/plots/xray.html} 
and the light curve of the GLE. 

We have verified from  literature that there is reported for most of the cases of only
GLE linked with energetic solar flares, those with a X-ray   
prompt emission classified as X-class (above $10^{-4} Watts\;m^{-2}$). These observations of solar   
flares \citep{smart96} 
have led to the identification of two classes of acceleration events: impulsive (prompt)  
and gradual(post-eruptive or delayed). The impulsive events require selective acceleration   
such as the gyroresonant interaction with plasma waves. The energetic particles  
from these events arrive very quickly, around 15 minutes after a flare. In contrast, the gradual   
events have a strong association with coronal mass ejection (CME) and suggest that the particles   
in these events are accelerated by CME driven shocks. The energetic particles of these events  
are observed up to several hours after a flare. The effect on the interplanetary medium occurs
preferentially during this post-eruptive phase \citep{kocharov95}.
In the case of the first GLE, the association is with the two last flares on December 02 2004, 
whose X-ray prompt  
emission (the first) is of low intensity (C5.0 class) and (the second) is of medium intensity   
(M1.5 class). The delay between the X-ray prompt emission and the GLEs is 1.5 hours. Consequently 
they can be associated with the gradual or post-eruptive events.
We would like to point out that the temporal shift between the two prompt X-ray emissions and the 
GLE's peaks is practically the same.

Besides the GOES report of X-ray prompt and proton emissions, there is also the report by
ACE EPAM WART60 of energetic nuclei
such as flows of nuclei ($E>100MeV$) of hydrogen, helium, oxygen and iron
emitted by the sun during the flares  in the proportion (at one AU from the SUN) 
$\sim 5000:100:10:10$ in units of $(cm^2,s,sr,MeV/nuc)^{-1}$ 
as hourly average flux \citep{shofield03}\footnote{(Shofield 2003)is available at
http://sd-www.jhuapl.edu/ACE/EPAM/janice/epmwww.cgi?2320+current+e}.
The beginning of the GLE coincides approximately with the decline of the Sun,
when the Sun, was not far from field of view of the TUPI telescope (see Table 2).

On the other hand, an accurate determination of the fluence of solar energetic particles from 
the flare that might generate muons  requires the obtaining of the muon energy spectrum. 
As already has been indicated, in the 
present stage of our experiment, the detector is only a directional counter telescope of muons 
above a energy threshold (~0.3 GeV). 

We have examined also the light curve for this GLE for other pulse-hight amplitude 
discrimination, as shown in Fig.5. The signal persists even when a high pulse amplitude is 
used as discrimination level.
We would like to comment that there is a third solar X-ray flare ($M1.4-class$) on 2003 December 2,
with a beginning at $12^h47^m$ UT. We did not register this because the raster scan only began 
at $13^h31^m$ UT

\subsection{The second ground level event (GLE 2003/12/16)}

The GLE 2003/12/16 was also detected during an off-source raster scan.   
The light curve shape is like a FRED (fast rise exponential decay)
as is shown in of Fig.6 (low panel).
The excesses for this FRED GLE can be seen over a 
background rate of approximately 18 coincidences per 10 seconds with significance levels of
$7.9\sigma$. The origin of this GLE is unknown, because  
there is neither a satellites report of a solar flare, nor of prompt X-ray emission, 
nor an excess of protons or heavy particles
during the raster scan where the GLE has been observed.
We would like to highlight  that the GLE's duration
is 416 seconds. This means that the GLE is not the remains of showers produced by
conventional cosmic rays.

The enhancement of muons at sea level from solar flares is linked with the most energetic
particles (i.e. protons, alphas...), those with energies above the pion production threshold
and above the cut-off rigidity of the region where the detector is located.
In fact, the pitch angle (where zero degree pitch angle represents sunward IMF direction)
of high energy particles excess linked with solar flare presents a large anisotropy distribution
\citep{duldig01},
with a systematic intensity excess around zero degree pitch angle and decreases 
up to ~60 degrees. The excess at large pitch angle, 
those up to 180 degree as observed by large field of view detectors such as neutron monitors,
are constituted by low energy particles. They are in the tail of the distribution, where the 
intensity is minimum.

Even so, the FRED GLE could be an event connected  via the file magnetic lines, with a flare on the other 
side of the Sun, and not seen by satellites. The probability of detection using a directional 
telescope of small field of view  (0.082 sr of angular window) pointed in a random direction, an event 
above the horizon is  approximately $p \sim 0.082/2\pi =0.013$. However, because the  telescope 
can detect a fraction of muons, $\Delta_{\mu}(r)$, even  when the core of the air shower is at a 
distance $r(\cong 2\;km)$ from the 
telescope center, the probability is enhanced to $\sim 3\%$ ($\sim 5\%$ for primary gamma-rays).
$\Delta_{\mu}(r)$ is calculated using the lateral distribution function of muons (see sec.6).
The detection of an event that happens on the other side of the Sun would correspond to a very large
pitch angle, and from the considerations mentioned above, we estimate the probability of detecting a solar 
flare connected to the back of the Sun at less than 4\%.

Another possibility (although remote) to explain the origin of this GLE 
is to invoke  the GRB hypothesis. 
The temporal and directional coincidences of a GLE with satellite observations of GRBs
are strong indications of a common detection. In fact this has been the main objective 
of several ground experiments, not only the detection of the GRBs TeV counterpart but 
also their afterglows at X-ray, optical and radio wavelengths. 
Nowadays it is expected that the rate of observation of GRBs by 
the GRB coordinate network (GCN) satellites 
\citep{barthelmy01}\footnote{(Barthelmy 2001) is available at http://gcn.gsfc.nasa.gov/$\#$tc4}
with the field of view of a large ground based detector such as MILAGRO is smaller
than one per month \citep{smith01}. 

However, we have found GRB satellite notification around two hours from the beginning of the
FRED GLE (see Table 2). While we don't have other evidences which indicates a common detection
with GRBs from GCN satellites. 
Fig.6 shows a comparison between the  
light curve shapes of the BATSE burst Trigger 7989 and the GLE 2003/12/16. Both are genuine FREDs.
We have also examined the light curve for this  GLE for other pulse-hight amplitude  
discrimination levels, as  shown in Fig.7.
The signal persists even when a high pulse amplitude is used as discrimination 
level, while the background is basically eliminated. 
It is possible to see that the signal observed in the GLE linked with the solar flare is more intense  
when compared with the signal of the FRED GLE.

On the other hand, there is evidence for two classes of bursts when they are classified 
according to  their duration. The result comes from BATSE catalog
\citep{paciesas99} as  shown in Fig.8. We can see   
that the duration distribution of 1234 BATSE bursts is bimodal. 
The first is centered around the a small value of $T90=0.2\;to\;0.3$ seconds and the 
second is centered around the a large value $T90=40\;to\;60$ seconds,
where $T90$ is the duration for 90\% of the bursts to occur. 
The arrow in Fig.8 shows the duration of the FRED GLE. We can see that   
its duration (416 seconds) is still inside that of the   
BATSE $T90$ distribution.

\section{Local effects}

The counting rate of scintillator detectors at ground level is subject
to several sources of modulation. The main ones are due to atmospheric pressure 
variation, solar activity and the 24 hours sidereal anisotropy. However, 
the temporal scales of
these modulation phenomena are much larger than the GLEs duration.

In the case of a tracking telescope like the TUPI it is necessary to take 
into account the geomagnetic effect responsible for the muon flux dependence
on azimuth angle (see section 3). Again the temporal scale to observe this effect 
is much larger than the GLEs duration.

In order to take into account possible anomalous pressure variations as being 
responsible of the GLEs, we have monitored the barometric pressure and
included this in our data acquisition system. Every 10 seconds
the counting rate and the atmospheric pressure are registered.
Under normal conditions, the daily (24 h) variations of the atmospheric pressure 
present a maximum value and a minimum value. This tendency has been found
during the raster scan where GLEs have been found. Fig.9 summarizes the 
situation where the pressure time series on 2003/12/02 is shown in the 
upper panel and its corresponding fast Fourier transformation (FFT) is shown in the 
lower panel. The arrow in the upper figure indicates the beginning of the GLE. 
The absence of peaks in the power spectrum means that there are no  
scintillation phenomena as indications of anomalies. 
The power spectrum gives an estimate  of the mean square fluctuations at frequency 
f and, consequently, of the variations over a time scale of order $1/f$.  
For the pressure case, the spectral density varies as $1/f^{1.96}$ and this 
quite steep spectral density is close to a correlated Brownian noise with $1/f^2$, 
over many decades, or in other words, over all the 12 hours of the raster scan. 

Consequently, pressure variation could not be a cause of the origin of the first GLE.
A similar situation has been found for the second GLE. 

\section{Is there an enhancement of muons at sea level from GRBs?}

There are several reports of ground level observations, of solar flares, especially   
those of high intensity \citep{bieber02,falcone99,swinson90}, as well as 
the enhancement of muons at ground level from solar  
flares of large scale has been reported \citep{poirier02,munakata01}.  
  However, the enhancement of muons at ground from GRBs remains open, at least  
they have not been observed with  high confidence level.  
The Milagrito experiment, a predecessor 
of the (water Cherenkov) Milagro experiment, has reported evidence for the TeV 
counterpart of a BATSE GRB
(970417) \citep{atkins00}. The GRAND project (muon detectors array at ground level) has 
reported some evidence of GRB detections in coincidence with one BATSE GRB (971110)
\citep{poirier03}, even if with a low significance.

Besides the experimental evidence, several models for the origin of GRBs predict 
a TeV component \citep{totani99,dermer00,pilla98}.
A plausible explanation for the extremely low rate of events observed in the TeV region
is to invoke the attenuation of TeV photons by interaction with the intergalactic 
infrared radiation. This process would be responsible for a cut off for TeV GRBs
whose sources are at distances larger than $z \sim 0.3$.
However, for GRBs with $E_{\gamma}> 20$ GeV, the cut off can be shifted to $z\sim 2$.
Consequently, in order to increase  the GRBs rate detected at ground level a search 
for GRBs in a energy region above $\sim 20$ GeV has to be undertaken.  

A primary photon converts to an $e^+e^-$ pair after 1 radiation length (on average),
$\lambda_R\sim 37\;g/cm^2$ in the atmosphere. In a subsequent radiation length
the electromagnetic particles further lose energy by bremsstrahlung 
$e^{\pm}\rightarrow \gamma e^{\pm}$. The processes are repeated forming an
electromagnetic air shower. All shower gamma rays (including the primary)
above the photo-production cross section can contribute to the muon content
in the shower. Despite the fact that gamma-rays 
in the energy band of 30 GeV to several TeV have only a 0.5\% to 3\% chance of 
undergoing interactions in the atmosphere yielding pions, the photons are more
efficient at producing energetic, forward directed pions. The result comes
from the Feynman-x distribution of charged pions obtained by using the FLUKA 
Monte Carlo simulation \citep{fasso01}. The $\gamma-Air$ interaction has a large fraction 
on high-x secondaries (pions) than $p-Air$ interaction. The FLUKA results 
also show that the distribution of height above sea level at which the detected muons are 
produced has a peak at $\sim 20\;km$  for proton showers \citep{batistone98} and 
$\sim 12\;km$  for photon showers \citep{poirier01}. 
In addition the distribution of the generation number of the grandparent
of the sea level muon in gamma showers is a very narrow distribution and is peaked at 
generation one for energies below a few hundred where the parent is mainly produced
in a photo-production process of the primary photon.
This means that "photo-muons" have a good chance of reaching ground level.

In order to estimate the number of muons reaching detection level which are
initiated by primary gamma ray interactions in the atmosphere, we have used
the muon kinetic energy spectrum within of a radial distance $R(=10\;km)$, from  
the shower center, on the basis of the FLUKA Monte Carlo program
\citep{fasso01}. For simplicity only vertical incidence of primary gamma rays is 
taken into account. The spectrum per incident photon can by parameterized
as
\begin{equation}
\frac{dN_{\mu}(>E_{\mu})}{dE_{\gamma}}\sim \frac{N_{\mu}^{0}}{A(R)}(\frac{E_{\gamma}}%
{GeV})^{\alpha}(\frac{E_{\mu}}{GeV})^{-\beta.},
\end{equation}
with all energies in GeV units. The parameters are
$N_{\mu}^{0}\sim 3.1\times 10^{-4}$, $\alpha\sim1.3$ and $\beta$ is close
to zero (integral flat distribution) in the muon energy band from 0.3 GeV to $\sim 20$
GeV and $A(R)=\pi \times R^2$. 
A very small fraction of these
muons are detected by the TUPI telescope with a acceptance,
$a_{\mu}$, which increases as the photon energy increases,
and is given by
\begin{equation}
a_{\mu}=a_{\mu}^{0}(\frac{E_{\gamma}}{GeV})^{\delta}\Delta(r),
\end{equation}
where $a_{\mu}^{0}\sim3.44\times10^{-5}$ and $\delta\sim 1.88$,
$\Delta (r)$ is the fraction of muons that hit the telescope when the core of 
the photon shower is at a distance r from the telescope center.
$\Delta (r)$ is calculated using the lateral muon distribution function.
According to the FLUKA results \citep{poirier01}, and for photon primary energies between
3 GeV to $10^4$ GeV, the lateral muon function distribution extend to more than 10 km,
whereas the sea level distribution have a relatively flat shape up to $r\sim 2$ km.
This means that $\Delta (r) \sim 1$ up to $r\sim 2$ km.

The biggest uncertainty of a GeV to TeV GRB is the shape of the primary photon
spectrum. We assume a GRB as constituted by $N_{\gamma}^{0}$ photons 
per $cm^{2}\;s\;GeV$,
arriving at the top of the atmosphere inside the field of view of the TUPI
telescope and with an energy spectrum of
\begin{equation}
\frac{dN_{\gamma}}{dE_{\gamma}}=N_{\gamma}^{0}(\frac{E_{\gamma}}%
{GeV})^{-\gamma},
\end{equation}
extending from $E_{min}$($\sim 1.0$ GeV) up to $E_{max}$ (several TeVs) and with a
duration $\Delta T$. The highest energy GRB observed by EGRET \citep{hurley94}  
suggests fluxes around $N_{\gamma}^0\sim 10^{-5}%
photons/cm^{2}s\;GeV$ and spectrum index around $\gamma \sim 2.0$, for energies
 near  $E_{\gamma} \sim 1.0$ GeV,. These values are used in the calculation.

Under these conditions, the number of muons (GeV muons) reaching the TUPI
telescope can be expressed as
\begin{equation}
N_{\mu}(>E_{\mu})\sim A_{eff} \times \Delta T \times a_{\mu}^{0}\times (N_{\mu}^{0}/A(R)) \times N_{\gamma}^{0}
\int_{E_{min}}^{E_{\max}}(\frac{E_{\gamma}}{GeV})^{\alpha-\gamma+\delta
}dE_{\gamma} \int_0^R \Delta (r) 2\pi dr.
\end{equation}
where $A_{eff}=Aper/\Delta \Omega \sim 5023\;cm^2$ is the effective area, 
 $Aper (=65.3 cm^2sr)$ is the aperture and $\Delta \Omega (=0.013\;sr)$ is the 
 angular window of the TUPI telescope.

The signal in the detector, $S=N_{\mu}(>E_{\mu})$, must be compared with the 
square root of noise, $\sqrt{N}$, given by
\begin{equation}
\sqrt{N}=\sqrt{I_{\mu} \times \Delta T \times Aper},
\end{equation}
where $I_{\mu}$ is the background muon intensity due to cosmic ray induced 
atmospheric showers.
The value of $S/\sqrt{N}$ necessary to consider an excess signal as a positive detection
is above 4. The sensitivity of TUPI muons for gamma bursts  is shown in Fig.10, for several
spectral indices in the power law energy spectrum and for a duration of the the burst of
$\Delta T (=100\;s)$. The observation of the bursts at ground level is
strongly limited by the spectral index value, as well as by the highest photon energies
of the spectrum. Bursts of very long duration, very high photon maximum energy  and with 
energy spectra not quite so steep can be observed at ground level by using telescopes with 
a small muon energy threshold ($\sim 0.3 GeV$).

The analysis on the basis of FLUKA's results shows a real possibility of  
observing under certain conditions a GRB by the TUPI telescope. 
A more accurate analysis, 
given the complexity of the processes, requires a full
Monte Carlo study including the detector response and geomagnetic effects on the 
charged muons.

\section{Conclusions}

We have reported a description and an analysis of two  GLEs observed
on December 2003 during a search for enhancements from the galactic center
with muons at sea level and detected by using the TUPI telescope, after an 
upgrade of the data acquisition system.  
The main conclusions are summarized as follows:

(a)The TUPI telescope can detect muons at sea level with energies greater than 
the $0.3$ GeV required to penetrate the two flagstones or walls surrounding 
the telescope. The concrete reduces 
the noise due to other non-muon particles, for example, it is opaque to electrons.
The telescope is sensitive to the conventional atmospheric muon flux. 
The muon flux obtained during a raster scan of 12 hours presents the well 
known West-East effect. In addition the geomagnetic effects distort 
the zenith angle distribution of sub-GeV to GeV muons

(b)The two GLEs analyzed here have been unambiguously detected, because 
in both cases the GLE signals survives even when
a large amplitude discrimination is used, while the background is 
eliminated. This means that the chance of then being
a fluctuation of the background is close to zero. In both cases, the GLEs 
are observed with a high significance level above $7.5\sigma$ (see Table2). 

(c)The observation of the two GLEs is not restricted to a simple excess. It is 
possible to see their temporal structures, such as the light curves. The first
GLE has a strong correlation with the two last solar flares on 2003/12/02.
While, the second GLE (the FRED GLE) has an unknown origin, because  
there is no satellite report of solar flares, nor of prompt X-ray emission, 
nor an excess of nuclei during the raster scan where the GLE was observed.
We estimate the probability of the origin of the FRED GLE be a solar flare   
connected to the back of the Sun as less than 4\%.

(d)The muon flux is subject
to several sources of modulation, such as the atmospheric pressure 
variation, solar activity and West-East effect, among others. However, 
the temporal scale of these modulation phenomena are much larger than 
the GLEs duration. In addition, no anomalous changes in the atmospheric pressure
have been observed, during the raster scan where the two GLEs were found. 

(e)We have found GRB satellite notification 52 minutes before the 
beginning of the FRED GLE (see Table2) and despite the FRED GLE duration being large (416 seconds), 
it is still inside the BATSE T90 duration distribution. As well as this, the long duration of 
the FRED GLE indicates that it is not the remains of showers produced by conventional cosmic rays. 
However, we don't have  other evidence which indicates a correlation with the GRBs detected 
by the GCN satellites.  

(f) Finally, we show, by using the FLUKA Monte Carlo 
results, that there is a window for observed GRBs constituted by photons with energies 
above 10 GeV via "photo-muon" production with energies at ground above $E_{\mu}\sim 0.3$ GeV. 
The enhancement of muons at sea level from primary gamma rays (for energies below 100 GeV)
it is a consequence of the lateral distribution function of muons at sea level 
in photo-showers being close to a flat distribution.
The fraction of muons that hit the telescope when the core of 
the photon shower is at a distance $r$ from the telescope center is the same for
$r$ up to $\sim 2$ km. In addition, for GRBs with flatter spectra, large duration and a high maximum 
photon energy the possibility of observation is enhanced (see Fig.10).

We are waiting for the next round of satellites with  large area telescopes such as the GLAST
\citep{atwood94}, they will be able to detect GRBs up to 300 GeV.
The GLAST will be able to confirm the estimates mentioned above.  
Until then, we conclude that the FRED GLE analyzed here is only 
a potential candidate for sub-TeV GRBs.

\acknowledgments

We are grateful to Dr. A. Ohsawa from Tokyo University for
help in the first stage of the experiment and to Dr. M. Olsen
for reading the manuscript. This work was partially supported by 
FAPERJ (Research Foundation of the State of Rio de Janeiro) in Brazil.

\clearpage


\begin{figure}
\plotone{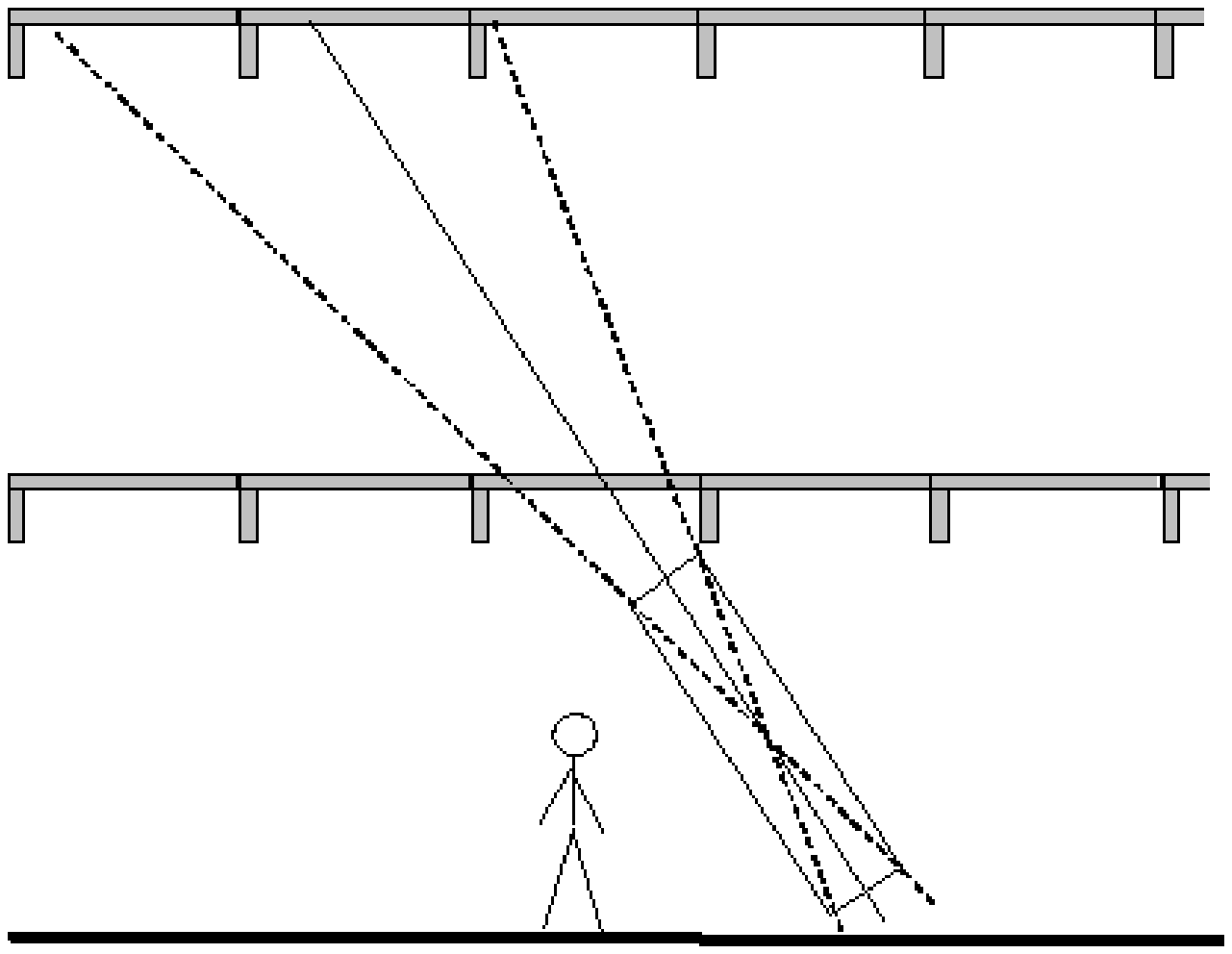}
\caption{General layout of the TUPI telescope, installed at sea level 
under two flagstones of concrete. \label{fig1}}
\end{figure}

\clearpage 

\begin{figure}
\epsscale{.80}
\plotone{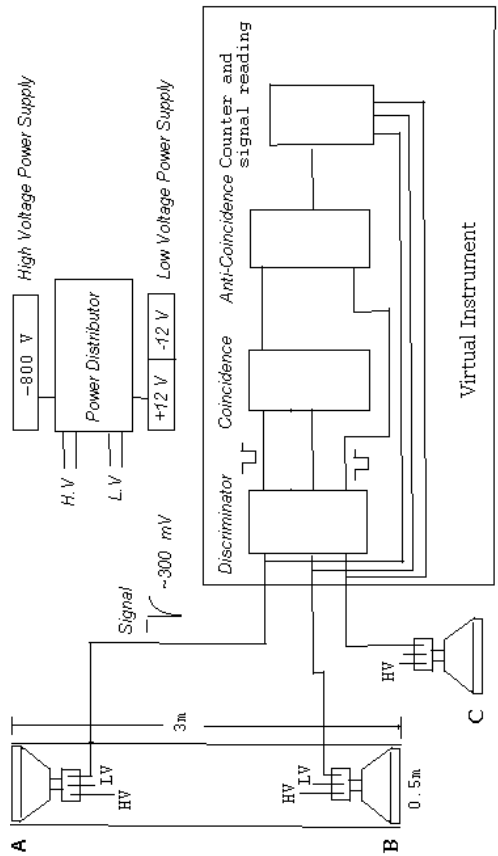}
\caption{TUPI data acquisition system: Block diagram.\label{fig2}}
\end{figure}

\begin{figure}
\epsscale{.75}
\plotone{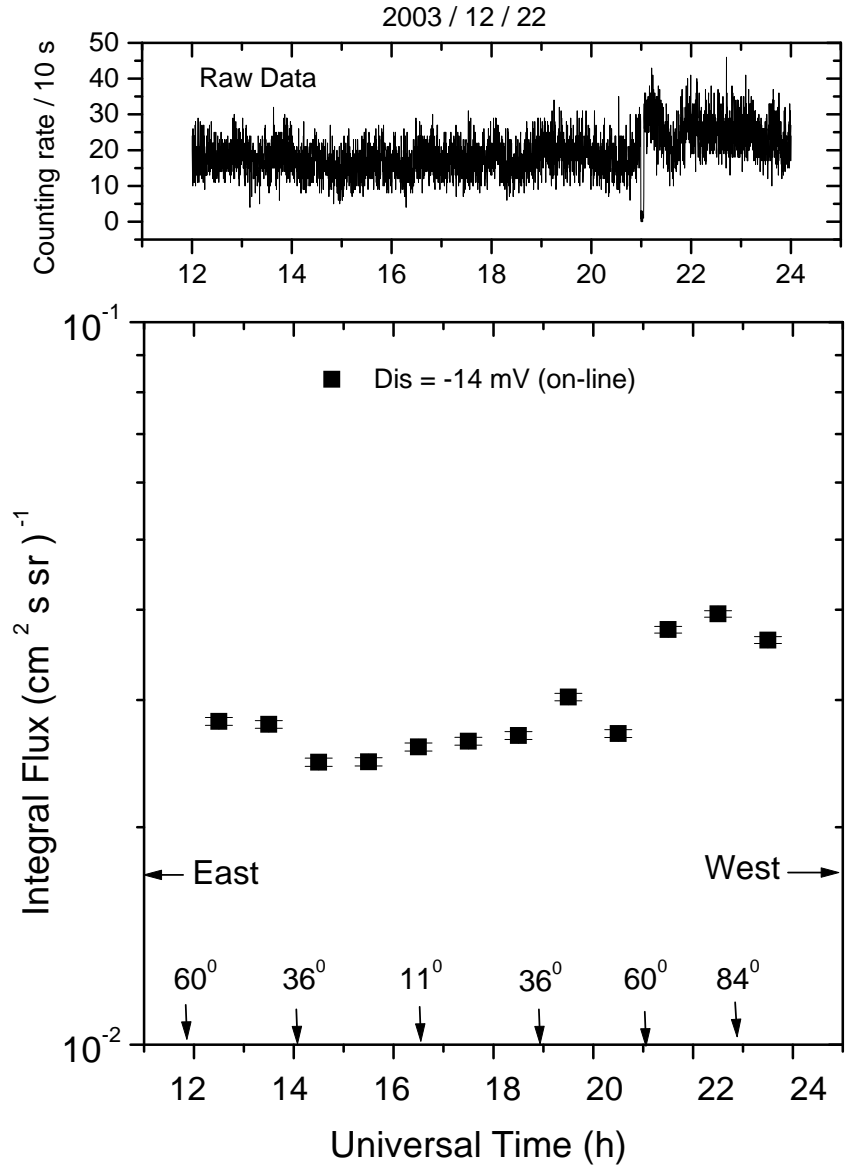}
\caption{Integral intensity of muons  
obtained during a typical raster scan. The zenith angle is indicated in 
the lower part.\label{fig3}}
\end{figure}

\begin{figure}
\epsscale{.75}
\plotone{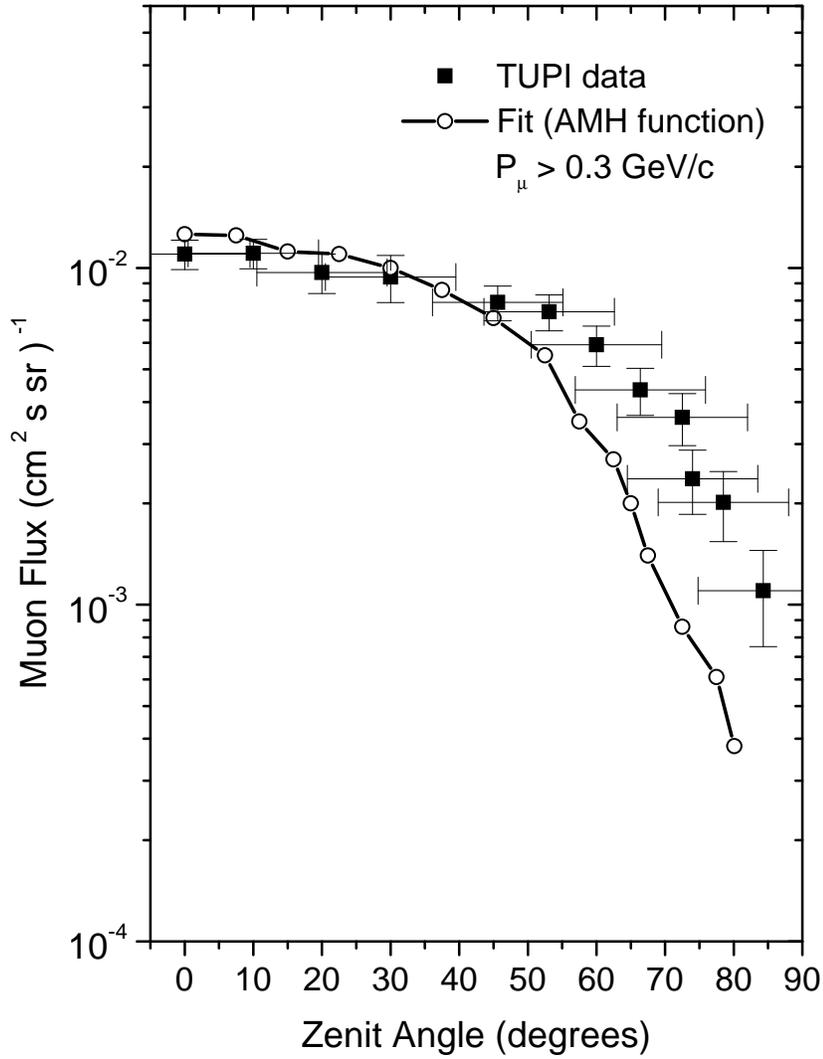}
\caption{Measured integral muon intensity, from $0^0$ up to $84.5^0$. 
The line represent a fit with the AMH (Texas A $\&$ M-University of Houston) function,
after a numerical integration. \label{fig4}}
\end{figure}

\begin{figure}
\epsscale{.80}
\plotone{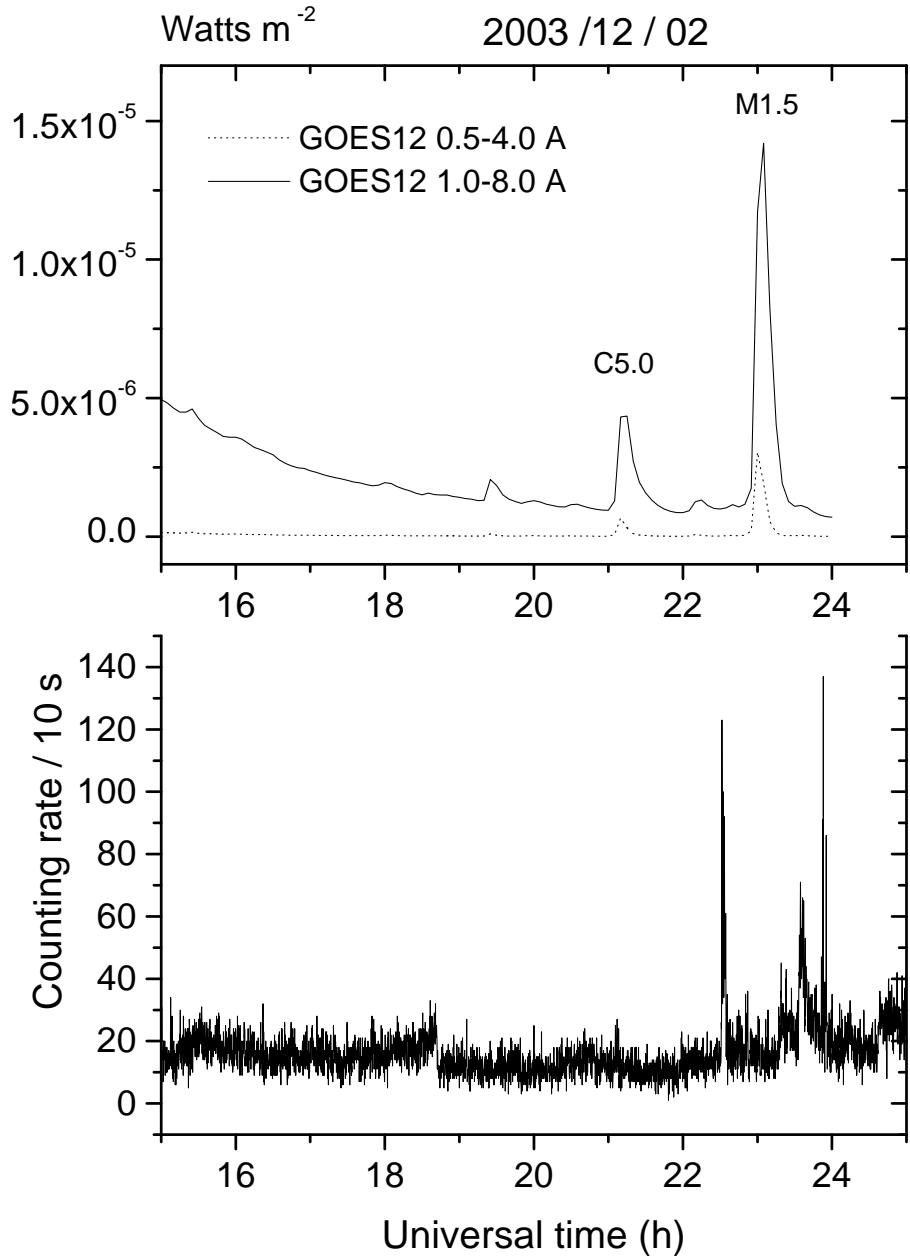}
\caption{Comparison between the  
light curve shapes of the GOES12 X-ray flux (flares) on 2003/12/02 and the GLE (2003/12/02).
\label{fig5}}
\end{figure}

\begin{figure}
\plotone{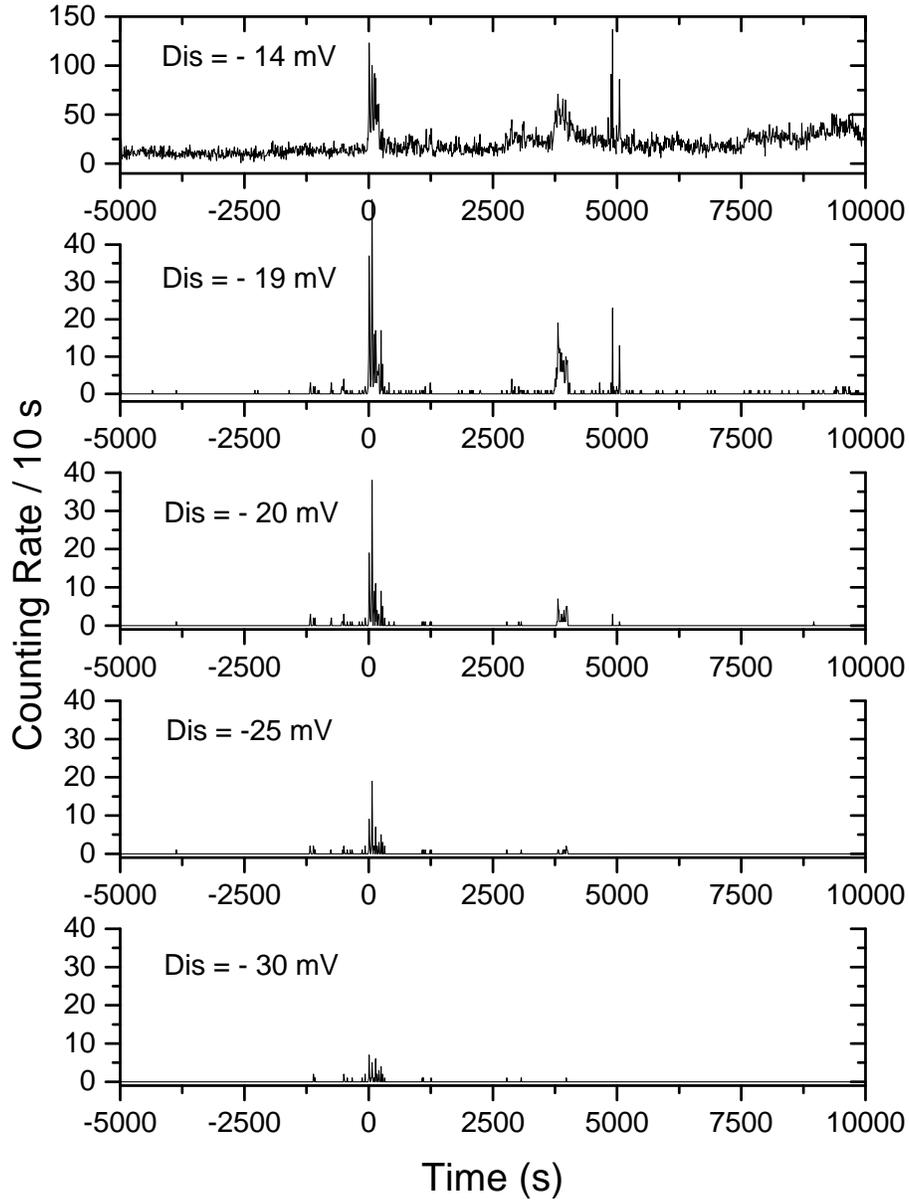}
\caption{Light curve shapes of the GLE (2003/12/02)
for different discrimination levels.
\label{fig6}}
\end{figure}

\begin{figure}
\epsscale{.75}
\plotone{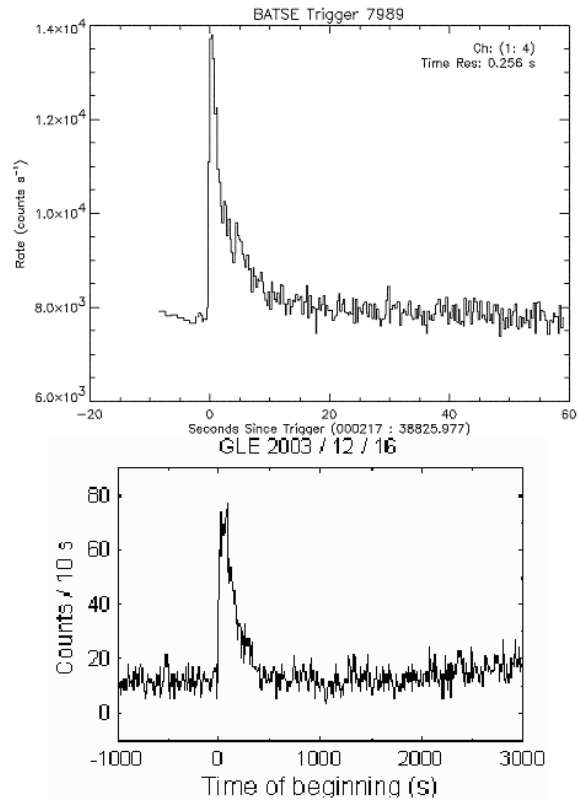}
\caption{Comparison between the  
light curve shapes of the BATSE/GRB010721 and the GLE (2003/12/16).
\label{fig7}}
\end{figure}

\begin{figure}
\epsscale{.75}
\plotone{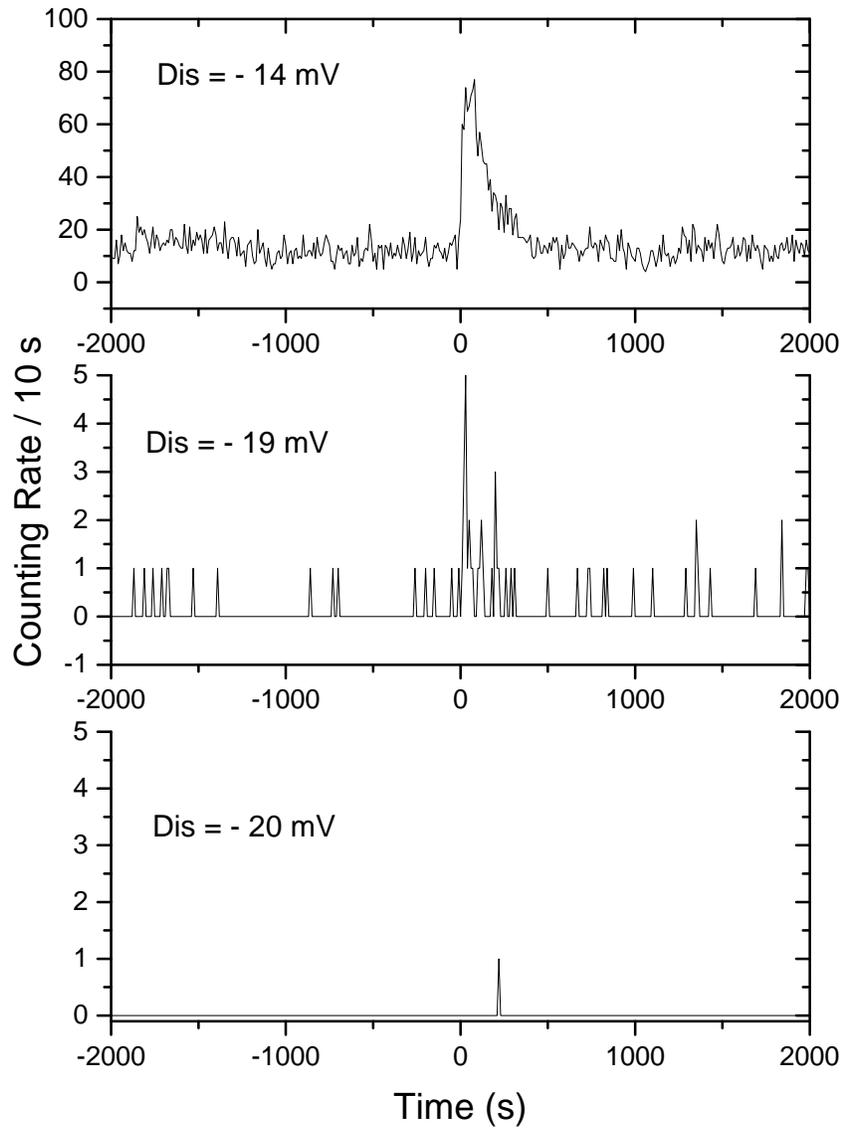}
\caption{Light curve shapes of the GLE (2003/12/16)
for three different discrimination levels.
\label{fig8}}
\end{figure}

\begin{figure}
\plotone{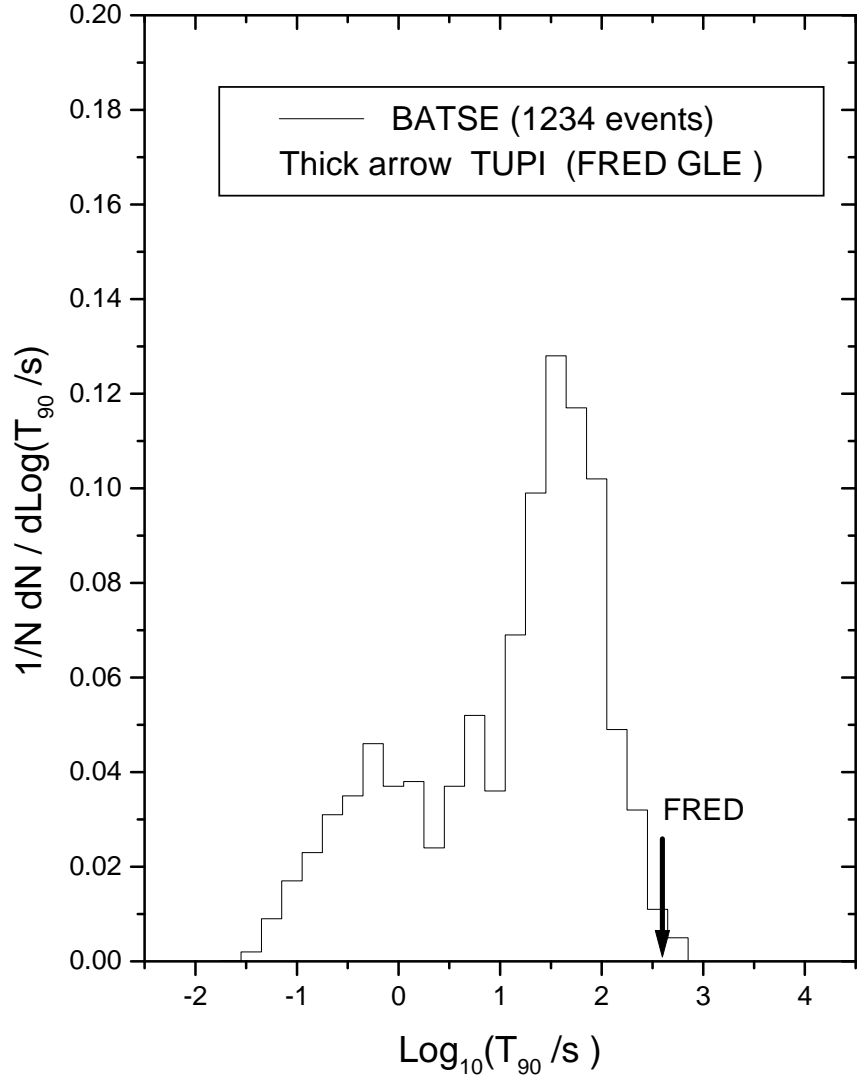}
\caption{Duration (T90) distribution of 1234 BATSE bursts.
The arrow shows the duration of the FRD GLE analyzed here. 
\label{fig9}}
\end{figure}

\begin{figure}
\epsscale{.75}
\plotone{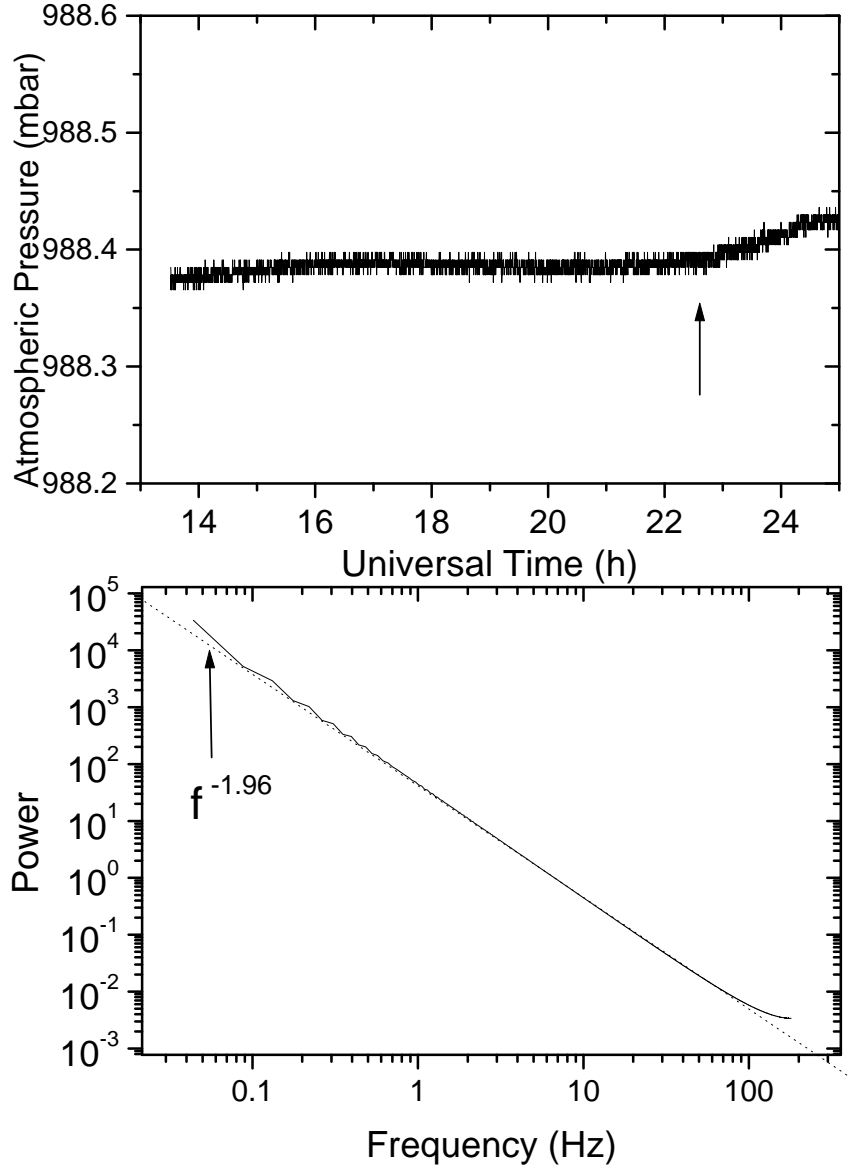}
\caption{Time series for the atmospheric pressure on 2003/12/02.
The arrow indicates the beginn0ing of the GLE (upper part), and its power spectrum
(lower part).
\label{fig10}}
\end{figure}

\begin{figure}
\plotone{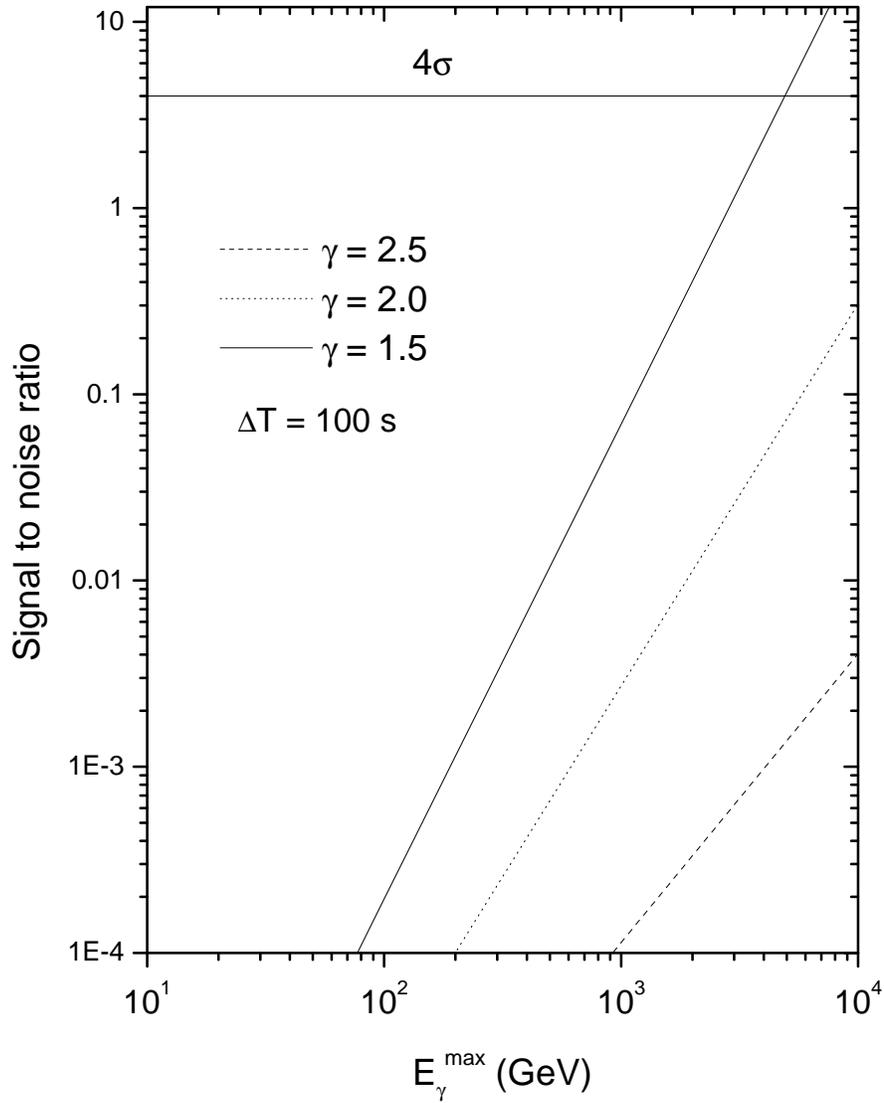}
\caption{Signal to Noise ratio due to a transient event of 100 s of duration
and several spectral index in the power law energy spectrum (see section 6).
According to photo-production FLUKA results and under the TUPI telescope 
conditions.
\label{fig11}}
\end{figure}

\clearpage

\begin{deluxetable}{crrrrrrrrrrr}
\tabletypesize{\scriptsize}
\tablecaption{West-East Asymmetry parameters. In the case of TUPI
experiment, the error is only statistical}
\tablewidth{0pt}
\tablehead{
\colhead{Experiment} & \colhead{Azimuth }   & \colhead{Zenith } &
\colhead{R. cutoff (GV)} &
\colhead{Energy (GeV)}  & \colhead{W-E Asymmetry}} 

\startdata
TUPI \tablenotemark{a} & South East-South West &$\theta=60^0$  & 9.8 &$> 0.3 $ & $(15\pm 3)\%$ \\ 
\hline
TUPI &       East-West       & $\theta=20^0$    & 9.8      &$> 0.3 $ & $(18\pm 3)\%$   \\    
Okayama& East-West &$\theta =20^0$ & 12             &$1-3   $ & $(17\pm 4)\%$ \\
\enddata

\tablenotetext{a}{In raster scan regime}



\end{deluxetable}

\begin{deluxetable}{crrrrrrrrrrr}
\tabletypesize{\scriptsize}
\tablecaption{Chronology and main characteristics of the two GLEs analyzed here.}
\tablewidth{0pt}
\tablehead{
\colhead{Data} & \colhead{Stard UT(h)}   & \colhead{Significance level}\tablenotemark{a}   &
\colhead{RA(deg)\tablenotemark{b}} &
\colhead{Dec(deg)\tablenotemark{b}}  & \colhead{Satellite Notification} & \colhead{RA(deg)} &
\colhead{Dec(deg)}     
}
\startdata
2003/12/02 & 22:31 &$10.5\sigma$ &295.5  &$-$29 &21:09(GOES12)Flare     & Sun(247.9) &Sun(-22) \\
2003/12/16 & 20:26 &$7.9\sigma$ &303.0 &$-$29 &19:34:22(HETE-2974)GRB &94.32 &$+$9.24 \\
\enddata


\tablenotetext{a}{At the peak}
\tablenotetext{b}{With reference to the telescope axis}


\end{deluxetable}


\end{document}